\documentclass[11pt,a4paper]{article}

\usepackage{geometry}
\usepackage{amssymb}
\usepackage{amsmath}

\newcommand{\brho}{\mbox{\boldmath $\rho$}}

\newcommand{\bF}{\mathbf{F}}

\begin{document}
	
	\begin{center}
		
		{\Large
		{\bf Response on the }
		
		{\it
		''Extended Comment on the Article 
		''Consistent Development of a Beam-To-Beam Contact Algorithm via the Curve to Solid Beam Contact - Analysis for the Non-Frictional Case''}
	
		Christoph Meier, Alexander Popp, Wolfgang A. Wall
		
		{\bf	published in ''Researchgate'' in October, 2018}
		}\vspace{11pt}
		
		{\large Alexander Konyukhov and Karl Schweizerhof}
		\vspace{11pt}
		
		{\small\itshape Karlsruhe Institute of Technology. Karlsruhe. Germany.}
	\end{center}

	A discussion about the existence and uniqueness of the Closest Point Projection (CPP) for surfaces as well as
	for the point-to-edge (curve)   situations has been first formulated for surfaces in the article of
	Konyukhov and Schweizerhof [1] and in the following article [2] for the curve-to-curve situation.
	This formulation is mathematically strict, because it is based on the well known theorem about
	the existence and uniqueness of the solution for the operator equation in which the operator is strictly convex.
	 As it has been proved in functional as well as in convex analysis, this statement
	 is valid not only for functions, but also for operators formulated in differentiable Hilbert spaces.
	 Application of this fact is widely found in different branches of physics including, of course, mechanics.
	 Thus, in Ogden [3] the uniqueness of the nonlinear elasticity problem has been discussed.
	 Namely this strict mathematical result has been employed by us to formulate the statement
	 for the existence and uniqueness of the CPP for all contact situations, see also details in monograph [4].
	 In the case of Curve-To-Curve (CTC) contact,
	 this criterion leads to the positivity for the second derivative of the distance function
	 $\bF(\xi^1, \xi^1) = \vert \vert \brho(\xi^1) -  \brho(\xi^2) \vert \vert $ between two curves $\brho(\xi^1)$
	 and $\brho(\xi^2)$. For further analysis the distance function
	 is formulated via angles $\varphi^1$ and $\varphi^2$,
	 determining the position of the osculating planes for two spatial curves
	 relatively to the corresponding Serret-Frenet coordinate systems attached to both curves.
	 The positiveness of the second derivative $\bF"$ is analyzed via
	  the Sylvester criteria for the positivity of the matrix.
		This statement is a system of nonlinear inequalities and, therefore, should be strictly  
		solved by using also mathematical methods for the solution of the system of inequalities.
		That is why in the aforementioned articles [2], we just geometrically interpreted 
		this statement using the projection domain, determining the position of osculating circles for both curves and the distance between them.
		In this domain the solution of the CPP procedure exists and is unique.
	
	In the article [5] our mathematical statement has been criticized and explained that a more general criterion has been found by them {\it (Citation:  “...our motivation to perform mathematically concise and rigorous investigation concerning the existence and uniqueness of the corresponding closest-point projection. On the contrary to Konyukhov et. al. who have already treated this question by means of geometrical criteria, we derive a very general analytical criterion that is valid for arbitrary contact formulations and that is based on proper and easy-to-determine control quantities.“)}.
	However,  the system of non-linear inequalities has not been solved and 
	even more restricted criterion, which is  valid only for a specific contact situation, has been developed!
	In order to show this, the following set of mathematically not consistent operations has been performed by authors in [5]:
	
\begin{enumerate}
	\item[1)]
	  the second inequality from the Sylvester criteria, namely $\det \bF" > 0$
	    has been estimated as ''the worst estimate''. 
	    The ''correctness'' of such an operation
	    will be obvious, if one can try to solve even the algebraic inequality $(x-a)(x-b) > 0 $ by ''the worst estimate'' method!
	    Any mathematically strict method for the inequality will require, of course,  consideration of the positivity for the function $\det \bF"$.  
	
		\item[2)] 		
	  during ''the worst estimate'' both angles $\varphi^1$ and $\varphi^2$ are taken such that
	    $\vert \cos \varphi_i  \vert = 1$. This operation, however,  geometrically leads to the enforcement of two osculating planes to coincide.
	
	 \item[3)]
	  finally, during the development of the  criterion and 
	  the worst estimation a ''radius of the beam $R$'' (?!) representing a circular cross-section of
	  a beam with a curve mid-line has been inserted.
	
\end{enumerate}

	By using these operations it is claimed that 
	{\it a very general analytical criterion for the uniqueness and existence}  has been derived. 
	This is definitely not the case! 
	It is rather an approximation of a solution for the criterion for straight thin beams resp. small curvature beams. 
	This solution involves a minimal angle between tangents $\psi \equiv \alpha_{crit} $
	at which a switch from Curve-To-Curve (CTC) to another algorithm Point-To-Curve (PTC) is performed. 
	This is an adaptive switch which was claimed to provide a smooth transition between PTC and CTC algorithms providing existence and uniqueness 
	for the CPP procedure. 

	 In the following article [6], we emphasized the fact, that a very general criterion by no means can be derived by performing ad-hoc and not consistent mathematical operations. First of all, none of the well known methods for solutions of the system of nonlinear inequalities has been employed, 
	 and,  second, the original problem must be formulated in differential geometry for the shortest distance between curves 
	 (which have no thickness (or radius of the beam $R$) at all!!!). 
	 In addition, already in the article [7] we emphasized, that the beam-to-beam contact algorithm can be based also on a Point-To-Curve (PTC) 
	 projection procedure. The latter leads to another more easier described projection domain only for one curve 
	 -- this algorithm has been first derived already in [1]. Both statements together with other cases have been widely discussed a while ago in our book [4]. Further scientific discussions among researchers, working in contact mechanics (including conferences), 
	 have been rising a question -- 
	 in which cases PTC or CTC algorithms are  applicable and how to switch between them. 
	 Thus, Litewka in [8] proposed ad-hoc criteria combining PTC and CTC together based on the angle between straight lines resp. straight linear finite elements.

	In our last article we ourselves still did not solve the system of inequalities in order to reach a general criterion for curves, 
	instead of this, 
	we have found out cases of multiple solutions for CPP for CTC which are fully independent from the introduced before critical angle 
	$\psi \equiv \alpha_{crit} $.
	Such cases are general parallel curves constructed so that the distance between
	them is constant $ \bF = const $,  and, therefore, the CTC contact algorithm, requiring both $\bF' \ne 0$ and
	$\bF" \ne 0$, is not applicable. The angle $\psi$ between tangents for curves can  be then arbitrary.
		Our  presented Curve-To-Surface formulation is just an alternative formulation for beam-to-beam contact with a general elliptical cross-section,
		 which -- however -- works in general for parallel curves  contrary to CTC. 
		 Only mathematically strict methods are involved in this formulation, in which we further do not find any of the, so-called, ''errors''
		 -- named in the extended comment of Meier et.al.
	In addition we also found out that the nature of
	switching between PTC and CTC lays on the full  3D contact Hertz formulation -- as a physical background -- rather than
	on angle $\psi$ between tangent lines as an algorithmic background, proving then the ad-hoc criteria already proposed by Litewka in [8].
	
	This and only this was a point our remark.  The ABC formulation as an algorithm for contact of slightly curved beams is not criticized. In fact the ABC formulation includes as other known algorithms, see e.g. Litewka [8], both CTC and PTC formulations and being programmed in one package will always work, because on secure side PTC will be switched on. We refer again to our article [7] and book [4], as we have also in our programs such a scheme.
	
	Summarizing our discussions, 
	in {\it the Extended Comment on the Article [9]}, that we received, 
	we again 
	did not find
	any general solution of CPP including any method of solution of inequalities either. 
	This is rather a confirmation of the specific nature of the contact criterion and the corresponding solution plus algorithmic scheme. 
	Surprisingly, the authors started to criticize our alternative Curve-To-Solid Beam 
	formulation, which is based on the CPP for surfaces and not based neither on CPP for PTC, nor on CPP for CTC as a closer look should have revealed.  
	While doing so, the aforementioned authors are apparently not fully following the logic in our paper. Thus, as an example, while criticizing our parallel curves, they claim in several places that it is not possible  to construct any parallel curves with an arbitrary  angle $\psi$  between tangent lines. However, we can propose a simple geometrical example without any mathematics: 
	\newline
We just have to remember that $\bF(\xi^1, \xi^2)$ is a distance function between two curves 
-- so one can easily wind a wire on a cylinder at any desired angle  $\psi$  -- in this case the distance between 
the axis  of a cylinder and the mid-curve of a wire will be constant $\bF=r=const$. 
Thus, any derivatives of $\bF$ are zero  (necessary condition of minimum $\bF'=0$ is fulfilled) 
and the criterion of uniqueness $\bF''=0$ is spoiled -- the solution is, of course, by construction multiple -- 
at any point the distance is constant, however the angle between the axis of a cylinder and the tangent line to a wire is definitely arbitrary.
Namely, this case is fully illustrating insensibility of the introduced in [5] criterion based on $\alpha_{crit}$~and~$\mu_{max}$.

	Beyond that we do not want neither comment the counter critics nor go to further debates, 
	but suggest to check the corresponding mathematical background in differential geometry [10].
	
	We are finally again emphasizing that the problem of existence and uniqueness of CPP for CTC is not yet fully resolved 
	and can be addressed to mathematicians experienced in differential geometry and numerical analysis.

\end{document}